\documentclass[aps,prd,preprint,groupedaddress]{revtex4-1}
\usepackage{graphicx}
\usepackage{dcolumn}
\usepackage{bm}
\usepackage{mathrsfs}
\usepackage{amsmath}
\usepackage{amssymb}
\usepackage{float}

\begin{document}


\title{Direct CP violation in multi-body $B$ decays with the $a^0_0(980)$--$f_0(980)$ mixing}

\author{Chao Wang}
\affiliation{Center for Ecological and Environmental Sciences, Key Laboratory for Space Bioscience $\&$ Biotechnology, Northwestern Polytechnical University, Xi'an 710072, China}

\author{Xian-Wei Kang}
\affiliation{College of Nuclear Science and Technology,
Beijing Normal University, Beijing 100875, China}

\author{Rui-Wu Wang}
\email[Corresponding author, Email: ]{wangrw@nwpu.edu.cn}
\affiliation{Center for Ecological and Environmental Sciences, Key Laboratory for Space Bioscience $\&$ Biotechnology, Northwestern Polytechnical University, Xi'an 710072, China}

\author{Xin-Heng Guo}
\email[Corresponding author, Email: ]{xhguo@bnu.edu.cn}
\affiliation{College of Nuclear Science and Technology,
Beijing Normal University, Beijing 100875, China}


\begin{abstract}
We predict that the $a_0^0(980)$-$f_0(980)$ mixing would lead to large CP violation. We calculate the localized direct CP asymmetry in the decays $B^\pm \to f_0(980) [a_0^0(980)] \pi^\pm \to \pi^+ \pi^- \pi^\pm $ via the $a_0^0(980)$-$f_0(980)$ mixing mechanism based on the hypothetical $q\bar q$ structures of $a^0_0(980)$ and $f_0(980)$ in the QCD factorization. It is shown that there is a peak for CP violation, which could be as large as 58\%, when the invariance mass of $\pi\pi$ is near the masses of $a^0_0(980)$ and $f_0(980)$. Since the CP asymmetry is sensitive to the $a_0^0(980)$-$f_0(980)$ mixing, measuring the CP violating parameter in the aforementioned decays could provide a new way to verify the existence of the $a_0^0(980)$-$f_0(980)$ mixing and be helpful in clarifying the configuration nature of the light scalar mesons.
\end{abstract}

\maketitle

CP violation is one of the most fundamental and important properties of the weak interactions. Even though it has been known since 1964 \cite{Christenson:1964fg}, we still do not know the source of CP violation completely. In the standard model, CP violation is originated from the weak phase in the Cabibbo-Kobayashi-Maskawa (CKM) matrix \cite{Cabibbo:1963yz,Kobayashi:1973fv}. Besides the weak phase, a large strong phase is also needed for the direct CP violation in decay processes. Usually, this large phase is provided by the short distance and long distance interactions. The short distance interactions are caused by QCD loop corrections, and the long distance interaction can be obtained by some phenomenological mechanisms, which is more sensitive to the structure of the final states. It was suggested long time ago that large CP violation should be observed in the $B$ meson systems. In the past few years, more attentions have been focused on CP violation in the multi-body $B$ meson decays both theoretically and experimentally. Inspired by the experimental progresses, more efforts should be carried out for precisely testing the Standard Model and looking for the new physics through CP violation in these decay processes.

The scalars below 1\,GeV play an important role in understanding nonperturbative QCD. The $a_0^0(980)$ and $f_0(980)$ mesons aroused considerable theoretical interests, but their structures are still controversial. These two mesons, with different isospin but the same spin parity quantum numbers and closed masses, lie near the threshold of the $K\bar K$ channel and both of them couple to $K\bar K$. Due to the fact that the amplitude of the isospin breaking transition is caused by the mass difference of $K \bar K$, a mixing will occur between the $f_0(980)$ and $a_0^0(980)$ intermediate states in the multi-body decays. The $a_0^0(980)$-$f_0(980)$ mixing was discovered theoretically in the late 1970s \cite{Achasov:1979xc} and has been studied experimentally in several decays by the BESIII collaboration recently \cite{Ablikim:2018pik}. However, the mixing mechanism between these two mesons is still short of firm experimental evidence. In previous works, it was found that the $\rho$-$\omega$ mixing, which is also introduced due to the isospin violation, generates large strong phases and thus enhances the CP violation when the invariant mass of the final $\pi\pi$ state is in the $\rho$-$\omega$ interference region. Inspired by the $\rho$-$\omega$ mixing, we expect the $a_0^0(980)$-$f_0(980)$ mixing may lead to large CP violation. Since the CP asymmetry contains more informations on the strong phase than the decay width, we propose to test the $a_0^0(980)$-$f_0(980)$ mixing by the much more subtle calculation of CP violation. We will investigate the direct CP violation in three-body decays of the $B$ meson and discuss the $a^0_0(980)$--$f_0(980)$ mixing.

\begin{figure}[bt]
\centering
\includegraphics[bb=71 593 551 720, width=0.8\textwidth]{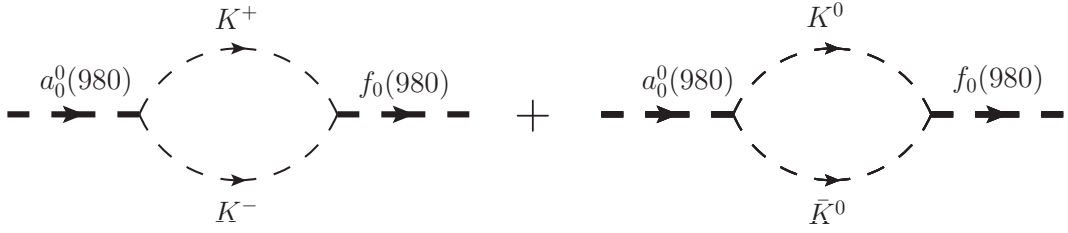}
\caption{The Feynman diagram for the $a^0_0(980)$-$f_0(980)$ mixing.}\label{a-f_mixing}
\end{figure}
Both $a^0_0(980)$ and $f_0(980)$ can couple to $K\bar K$. Due to the nonzero difference between the $K^+K^-$ and $K^0\bar K^0$ masses, the isospin break processes, $a^0_0(980)\rightarrow (K^+K^-+K^0\bar K^0) \rightarrow f_0(980)$ and $f_0(980)\rightarrow (K^+K^-+K^0\bar K^0) \rightarrow a^0_0(980)$, appear in the narrow region of the $K\bar K$ thresholds, which are shown in Fig.~\ref{a-f_mixing}. This $K\bar K$ loops would lead to an mixing amplitude. In Ref.~\cite{Achasov:1979xc}, this mixing mechanism was investigated phenomenologically. The amplitude of the $a_0^0(980)$-$f_0(980)$ mixing can be written as \cite{Achasov:1979xc,Achasov:2017zhu}
\begin{eqnarray}
\Pi_{a^0_0f_0}(m^2)&=&\frac{g_{a^0_0K^+K^-}g_{f_0K^+K^-
}}{16\pi}\Bigg\{\,\text{i}\,\Big[\rho_{K^+K^-}(m^2)
-\rho_{K^0\bar K^0}(m^2)\Big]-
\frac{\rho_{K^+K^-}(m^2)}{\pi}\ln\frac{1+\rho_{K^+K^-}(m^2)}
{1-\rho_{K^+K^-}(m^2)}\nonumber\\
&&\hspace{1cm}  +\frac{\rho_{K^0 \bar K^0}(m^2)}{\pi}\ln\frac{1+\rho_{K^0\bar K^0}(m^2)}{1-\rho_{K^0\bar
K^0}(m^2)}\,\,\Bigg\}\nonumber\\[5pt]
&\approx&\frac{g_{a^0_0K^+K^-}g_{f_0K^+K^-
}}{16\pi}\mathrm{i}\Big[\rho_{K^+K^-}(m^2)-\rho_{K^0\bar
K^0}(m^2)\Big],
\end{eqnarray}
where $g_{a^0_0(f_0)K^+K^-}$ are the effective coupling constants, $\rho_{K\bar K}(m^2)=\sqrt{1-4m_K^2/m^2}$ when $m$ (the invariant masses of scalar resonances) $\geq2m_{K}$, and $\rho_{K\bar K}(m^2)$ should be replaced by ${\mathrm i}|\rho_{K\bar K}(m^2)|$ in the region $0\leq m \leq2m_K$.

In recent years, the LHCb Collaboration has focused on multi-body final states in the decays of the $B$ mesons and preformed a novel strategy to probe CP asymmetries in the Dalitz plots \cite{Aaij:2013bla}. These multi-body decays provide much more information on strong phases than the two-body decays. Naturally, the $a_0^0(980)$-$f_0(980)$ mixing, if existing, would affect CP violation of the sequential three-body decay $B\to f_0(980)[a^0_0(980)] P\to \pi\pi P$ ($P$ represents a pseudoscalar meson) when the invariant mass of the final $\pi\pi$ locates around 980\,MeV.

For the sequential decays $B\to f_0(980)[a^0_0(980)] P\to \pi\pi P$, the decay width has the form
\begin{eqnarray}
\frac{\mathrm{d}\Gamma}{\mathrm{d}m}=\frac{m}{16 \pi^3 m_B^2}|{\bf p_1^* }| |{\bf p_3}| |\mathcal{M}|^2,
\end{eqnarray}
where $m$ is the invariant mass of $\pi\pi$, $m_B$ is the mass of the $B$ meson, ${\bf p_1^* }[=(1-4 m_\pi^2/m^2)^\frac{1}{2}]$, and ${\bf p_3}=[m_B^4-2m_B^2(m^2+m_P^2)+(m_B^2-m_P^2)^2]/(2m_B)$ with $m_P$ being the mass of $P$. The amplitudes can be expressed as
\begin{eqnarray}\label{r}
\mathcal{M}=\langle \pi^+\pi^- P|\mathcal{H}^T |B\rangle+\langle \pi^+\pi^- P|\mathcal{H}^P |B\rangle=\langle \pi^+\pi^- P|\mathcal{H}^T |B\rangle\left[1+r \mathrm e^{{\mathrm i}(\delta+\phi)}\right],
\end{eqnarray}
where $\mathcal{H}^T$ and $\mathcal{H}^P$ are the tree and penguin operators, respectively, we also define the strong phase $\delta$, the weak phase $\phi$, and the relative magnitude $r$, respectively. Considering the $a_0^0(980)$-$f_0(980)$ mixing, one has
\begin{eqnarray}\label{ama-f}
\langle \pi^+\pi^- P|\mathcal{H}^T |B\rangle&=&\frac{g_{f\pi\pi} T_{f_0}}{D_{f_0}} +\frac{ g_{f\pi\pi} T_{a^0_0} \Pi_{a^0_0f_0} }{D_{a_0^0}D_{f_0}-\Pi_{a^0_0f_0}^2},\nonumber\\[5pt]
\langle \pi^+\pi^- P|\mathcal{H}^P |B\rangle&=&\frac{g_{f\pi\pi} P_{f_0}}{D_{f_0}} +\frac{ g_{f\pi\pi} P_{a^0_0} \Pi_{a^0_0f_0} }{D_{a_0^0}D_{f_0}-\Pi_{a^0_0f_0}^2} ,
\end{eqnarray}
in which $T_{a^0_0(f_0)}$ and $P_{a^0_0(f_0)}$ correspond to the tree and penguin diagram amplitudes for $B\to a_0^0(980)[f_0(980)] P$, respectively, $g_{f\pi\pi}$ is the effect couple constant of $f_0(980)\to\pi^+\pi^-$, $D_{a_0^0}[D_{f_0}]$ is the inverse propagator of $a_0^0(980)[f_0(980)]$ constructed with taking into account the finite width. The expression of $D_{a_0^0(f_0)}$ can be written as
\begin{eqnarray}
D_{a_0^0(f_0)}(s)=m_{a_0^0(f_0)}^2-m^2+\sum_{ab}\left [\Re e {\Pi^{ab}_{a_0^0(f_0)}(m_{a_0^0(f_0)}^2)}- \Pi^{ab}_{a_0^0(f_0)}(m_{a_0^0(f_0)}^2)\right ],
\end{eqnarray}
where $\Pi^{ab}_{a_0^0(f_0)}$ are the diagonal matrix elements of the polarization operator corresponding to the contribution of the $ab$ intermediate states with $ab=(\pi\pi, K\bar K)$ for $f_0(980)$ and $ab=(\eta \pi, K\bar K)$ for $a^0_0(980)$.

Substituting Eq.~(\ref{ama-f}) into Eq.~(\ref{r}), we have
\begin{eqnarray}
r\mathrm{e}^{\mathrm{i}(\delta+\phi)}=\frac{\langle \pi^+\pi^- P|\mathcal{H}^P |B\rangle}{\langle \pi^+\pi^- P|\mathcal{H}^T |B\rangle} \thickapprox \frac{P_{f_0}D_{a_0^0} +P_{a^0_0} \Pi_{a^0_0f_0} }{T_{f_0} D_{a_0^0} +T_{a^0_0} \Pi_{a^0_0f_0}}.\label{rdelta}
\end{eqnarray}
It can be seen from Eq.~(\ref{rdelta}) that the $a^0_0(980)$-$f_0(980)$ mixing provides additional complex terms which may enlarge the CP-even phase and leads to a peak for CP violation when the invariance mass of $\pi\pi$ is near the $a_0^0(980)$ and $f_0(980)$ mesons. The differential CP violating parameter can be defined as
\begin{eqnarray}
A_{CP}\equiv\frac{|\mathcal{M}|^2-|\mathcal{\bar M}|^2}{|\mathcal{M}|^2+|\mathcal{\bar M}|^2}=\frac{-2r\sin \delta \sin \phi}{1+2r\cos \delta \cos \phi+r^2},\label{acpdefine}
\end{eqnarray}
where $m$ is near 980\,MeV in our case, $\mathcal{\bar M}$ is the decay amplitude of the CP conjugate process.

By defining
\begin{eqnarray}
\frac{P_{a^0_0}}{T_{f_0}}\equiv r^\prime \mathrm{e}^{\mathrm{i}(\delta_q+\phi)},\qquad \frac{T_{a^0_0}}{T_{f_0}}\equiv \alpha \mathrm{e}^{\mathrm{i}\delta_\alpha}, \qquad\frac{P_{a^0_0}}{P_{f_0}}\equiv \beta \mathrm{e}^{\mathrm{i}\delta_\beta},
\end{eqnarray}
to the leading order of $\Pi_{a_0^0f_0}$, we have
\begin{eqnarray}
r\mathrm{e}^{\mathrm{i}\delta} = \frac{r^\prime \mathrm{e}^{\mathrm{i}\delta_q} }{D_{a_0^0}} \left [\Pi_{a^0_0f_0} +\beta \mathrm{e}^{\mathrm{i}\delta_\beta} (D_{a_0^0}-\alpha \mathrm{e}^{\mathrm{i}\delta_\alpha}  )\right ].
\end{eqnarray}
$\delta_\alpha$, $\delta_\beta$ and $\delta_q$ denote the remaining unknown strong phases at short distance. In the absence of the $a^0_0(980)$-$f_0(980)$ mixing, at least one of these phases would have to be non-zero for CP asymmetry. In the ideal case, we neglect the strong phases at short distance. One has $r \mathrm{e}^{\mathrm{i}\delta}\propto \Pi_{a^0_0f_0}/D_{a^0_0}$. The isospin-breaking for $\rho$ and $\omega$ is generated at the quark level (between the $u$ and $d$ quarks) while the $a^0_0(980)$-$f_0(980)$ mixing is caused by the mass difference between $K^0 \bar K^0$ and $K^+ K^-$. Therefore, the modulus of $\Pi_{a^0_0f_0}$ ($|\Pi_{a^0_0f_0}|\approx 0.03\, \text{GeV}^2$ \cite{Achasov:2017zhu}) is larger than that of the $\rho$-$\omega$ mixing ($|\Pi_{\rho\omega}|\approx 0.0045\,\text{GeV}^2$ \cite{Wolfe:2010gf}), that is to say, the $a^0_0(980)$-$f_0(980)$ mixing can enlarge CP violation much more than the $\rho$-$\omega$ mixing. We also note that the phase of the mixing amplitude varies from 0 to $\pi/2$ when the invariant mass of $\pi^+\pi^-$ is near the 980\,MeV \cite{Achasov:2017zhu}. Thus, the CP violating parameter has a narrow peaks located around 980\,MeV in the final $\pi^+\pi^-$ spectrum.

Without loss of generality, we will now evaluate CP violation in the decays $B^\pm \to f_0(980) [a_0^0(980)] \pi^\pm \to \pi^+ \pi^- \pi^\pm $. Since we study the local CP violation in the narrow region near the masses of $f_0(980)$ and $a_0^0(980)$, the first step of the sequential decay $B^\pm \to f_0(980) [a_0^0(980)]\pi^\pm$ respects a simple factorization relation. We shall use the quasi two-body QCD factorization approach to deal with the amplitudes \cite{Cheng:2016ajl}. It is known that the underlying structures of the scalar mesons are not well established. The mesons $f_0(980)$ and $a_0^0(980)$ can be interpreted as conventional $q\bar q$ mesons, $q\bar q q\bar q$ multi-quark states or meson-meson bound states, and even scalar glueballs. If the scalar mesons are four-quark states, one more quark-antiquark pair is required to be produced in the decay process comparing with two quark picture. It is thus expected that the amplitude would be smaller in the four-quark picture than that in the two-quark picture when a light scalar meson is produced. Therefore, we assume that the $q\bar q$ structure dominates in our process and calculate the amplitude based on the $q\bar q$ structure.

In the QCD factorization, with the $q\bar q$ structure assumption, the amplitude of the decays $B^- \to f_0(980) [a_0^0(980)] \pi^-$ can be written as \cite{Cheng:2005nb,Cheng:2013fba}
\begin{eqnarray}
A(B^-\to f_0\pi^-)&=&
-\frac{G_F}{\sqrt{2}}\sum_{p=u,c}\lambda_p\Bigg\{ \left(a_1 \delta^p_u+a_4^p-r_\chi^\pi a_6^p+a_{10}^p-r_\chi^\pi a_8^p \right)_{f_0^u \pi} \nonumber \\
&\times& f_\pi F_0^{Bf_0^u}(m_\pi^2)(m_B^2-m_{f_0}^2)+\left(a_6^p-{1\over 2}a_8^p\right)_{\pi f_0^u}\bar r_\chi^{f_0}\bar f^u_{f_0}\,F_0^{B\pi}(m^2_{f_0})(m_B^2-m_\pi^2)\nonumber \\
&-& f_B\bigg[\big(b_2\delta_u^p+b_3+b_{\rm 3,EW}\big)_{f_0^u \pi} +\big(b_2\delta_u^p+b_3+b_{\rm 3,EW}\big)_{\pi f_0^u}\bigg] \Bigg\},
\end{eqnarray}
\begin{eqnarray}
A(B^- \to a^0_0\pi^- )&=&-\frac{G_F}{2}\sum_{p=u,c}\lambda_p\Bigg\{ \left( a_1\delta_u^p+ a_4^p-r_\chi^\pi a_6^p
 +a_{10}^p-r_\chi^\pi a_8^p \right)_{a_0\pi} \nonumber \\
&\times& f_\pi F_0^{Ba_0}(m_\pi^2)(m_B^2-m_{a_0}^2)-\left(a_6^p-{1\over 2}a_8^p\right)_{\pi a_0}\bar r_\chi^{a_0}\bar
 f_{a_0} (m_B^2-m_\pi^2)F_0^{B\pi}(m_{a_0}^2) \nonumber \\
&-& f_B\Big[\big(b_2\delta_\mu^p+b_3+b_{\rm 3,EW}\big)_{a_0\pi}-\big(b_2\delta_\mu^p+b_3 +b_{\rm 3,EW}\big)_{\pi a_0} \Big] \Bigg\},
\end{eqnarray}
where $\lambda_p=V_{pb}V_{pd}^*$, $f_{B,\,\pi,\,f_0,\,a_0}$ and $m_{\pi,\,f_0,\,a_0}$ are the decay constants and the masses of the corresponding mesons, respectively, $\bar f_{S}=f_{S} \frac{2m_\pi^2}{m_b(\mu)(m_u(\mu)+m_s(\mu))}$ ($\mu$ is the scale parameter), $F_0^{Bf_0^u,\, Ba_0,\,B\pi}$ are form factors, and
\begin{eqnarray}
\bar r_\chi^{a_0}=\frac{2m_{a_0}}{m_b(\mu)},\qquad \bar r_\chi^{f_0}=\frac{2m_{f_0}}{m_b(\mu)}, \qquad r_\chi^\pi=\frac{2m_\pi^2}{m_b(\mu)(m_u(\mu)+m_s(\mu))}.
\end{eqnarray}
The general form of the coefficients $a_i^p(M_1M_2)$ at the next-to-leading order in $\alpha_s$ are
\begin{eqnarray}
a_i^p(M_1\,M_2)=c_i+\frac{c_{i\pm1}}{N_c}+\frac{c_{i\pm1}}{N_c}\frac{C_F\alpha_s}{4\pi}\Big[V_i(M_2)+\frac{4\pi^2}{N_c}H_i(M_1\,M_2)+P^p_i(M_2)\Big],
\end{eqnarray}
where $c_i$ are the Wilson coefficients, the upper (lower) sign of $\pm$ corresponds to $i$ being odd (even), $C_F=(N_c^2-1)/(2N_c)$ with $N_c=3$, $M_2$ is the emitted meson and $M_1$ shares the same spectator quark with the $B$ meson, $V_i(M_2)$, $P_i(M_2)$, and $H_i(M_1,\,M_2)$ are vertex corrections, penguin corrections, and hard spectator corrections, respectively, which are listed in Appendix. As for the weak annihilation contributions, we follow the expressions given in Refs.~\cite{Cheng:2005nb,Cheng:2013fba}
\begin{eqnarray}
b_1&=&\frac{C_F}{N_c^2}c_1A^i_1, \qquad \qquad b_1=\frac{C_F}{N_c^2}\left[c_3A^i_1+c_5(A^i_3+A^f_3)+N_cc_6A^f_3 \right],\nonumber\\
b_1&=&\frac{C_F}{N_c^2}c_2A^i_1, \qquad \qquad b_1=\frac{C_F}{N_c^2}\left[c_4A^i_1+c_6A^f_2\right],\nonumber\\
b_{3,\mathrm{EW}}&=&\frac{C_F}{N_c^2}\left[c_9A^i_1+c_7(A^i_3+A^f_3)+N_cc_8A^f_3 \right],\quad b_{4,\mathrm{EW}}=\frac{C_F}{N_c^2}\left[c_{10}A^i_1+c_8A_2^i\right],
\end{eqnarray}
where the subscripts 1, 2, 3 of $A_n^{i,\,f}$ denote the annihilation amplitudes induced from $(V-A)(V-A)$, $(V-A)(V+A)$ and $(S-P)(S+P)$ operators, respectively, and $i$ and $f$ denote the gluon emission from the initial- and final-state quarks, respectively.

In the naive $q\bar q$ model, $f_0(980)$ is an pure $s\bar s$ state. However, several experiments indicate that there is a mixture between the light and strange quarks:
\begin{eqnarray}
|f_0(980)\rangle&=&|s\bar s\rangle \cos{\theta} + |n \bar n  \rangle\sin{\theta},\nonumber\\
|f_0(500)\rangle&=&-|s\bar s\rangle \sin{\theta} + |n \bar n  \rangle\cos{\theta},\nonumber
\end{eqnarray}
where $n \bar n \equiv (u \bar u +d\bar d)/\sqrt{2}$ and $\theta$ is the mixing angle. We take $\theta=20^{\circ}$ as in Refs.~\cite{Cheng:2005nb,Cheng:2013fba}. In the expressions of the spectator and annihilation corrections, there are end-point divergences $X_{H(A)}\equiv \int^1_0\mathrm d x/(1-x)$. The QCD factorization suffers from these endpoint divergences which can be parameterized as \cite{Cheng:2005nb,Cheng:2013fba}
\begin{eqnarray}
X_{H(A)}=\ln{\frac{m_B}{\Lambda_h}}\left[1+\rho_{H(A)}\mathrm{e}^{\mathrm{i}\phi_{H(A)}}\right],
\end{eqnarray}
where one may take $\rho_{H(A)}\leq 0.5$ and arbitrary strong phases $\phi_{H(A)}$ \cite{Cheng:2005nb}. In practice, we calculate the CP violation when $\rho_{H(A)}=0.25,\,0.5$ and $\phi_{H(A)}=0, \pi/2, \pi, 3\pi/2$, respectively.

\begin{figure}[bt]
\centering
\includegraphics[bb=2 285 583 661, width=0.8\textwidth]{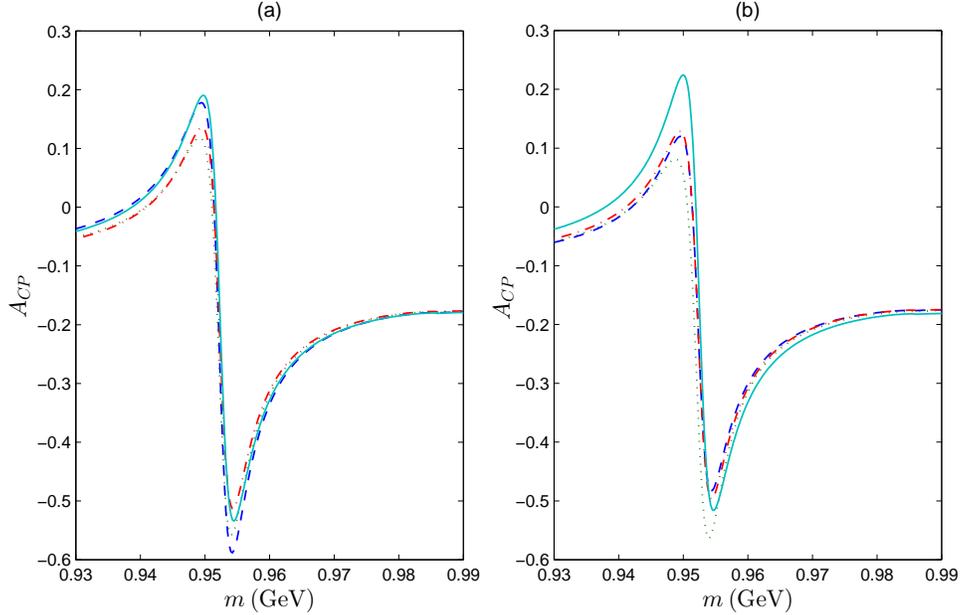}
\caption{The differential CP violating parameter as a function of $m$ in the decays $B^\pm \to f_0(980) [a_0^0(980)] \pi^\pm \to \pi^+ \pi^- \pi^\pm $. {\bf a} For $\rho_{H(A)}=0.25$: the dash line, dot line, dot-dash line and solid line correspond to the cases when $\phi_{H(A)}=0, \pi/2, \pi, 3\pi/2$, respectively. {\bf b} The same as {\bf a} but for $\rho_{H(A)}=0.5.$}\label{num}
\end{figure}

Then, considering the $a^0_0(980)$--$f_0(980)$ mixing, the total amplitude of the decay $B^- \to f_0(980) [a_0^0(980)]\pi^- \to\pi^+ \pi ^-\pi^-$ can be written as
\begin{eqnarray}
\mathcal{M}(B^-\to \pi^+ \pi^-\pi^-)&=&\frac{g_{f\pi\pi}}{D_{f_0}(m)}A(B^- \to f_0(980) \pi^-)\nonumber\\
&&\qquad +\frac{ g_{f\pi\pi}\Pi_{a^0_0f_0} (m)}{D_{a_0^0}(m)D_{f_0}(m)-\Pi_{a^0_0f_0}^2(m)} A(B^- \to a_0^0(980) \pi^-),
\end{eqnarray}
and the differential CP violating parameter is
\begin{eqnarray}
A_{CP}=\frac{|\mathcal{M}(B^-\to \pi^+ \pi ^-\pi^-)|^2-|\mathcal{M}(B^+\to \pi^+ \pi ^-\pi^+)|^2}{|\mathcal{M}(B^-\to \pi^+ \pi ^-\pi^-)|^2+|\mathcal{M}(B^+\to \pi^+ \pi ^-\pi^+)|^2}.
\end{eqnarray}
One can see that $\lambda_p(\lambda^*_p)$ provides a weak CP-violation phase and the strong phases occur in the mixing amplitude, the scalar propagators, and $V_i(M_2)$, $P_i(M_2)$ and $H_i(M_1,\,M_2)$ in the decays $B^\pm\to f_0(980) [a_0^0(980)]\pi^\pm$. Since we focus on the local CP violation in the narrow region near 980\,MeV, it is plausible to assume that the intermediate states $f_0(980)$ and $a_0^0(980)$ are on shell. According to the kinematics of the sequential decay, the amplitudes and the CP violating parameter are dependent on the invariant mass of $\pi^+\pi^-$, $m$. Substituting Eqs.~(10), (11), (13), (14) and (16) into Eq.~(17), the differential CP violating parameter is obtained, which is displayed in Fig.~\ref{num} as a function $m$. In the effective region of the $a_0^0(980)$-$f_0(980)$ mixing, we can see that the CP violating parameter can reach as large as 58 percent and an obvious peak exists, which can be used to test the existence of the $a_0^0(980)$-$f_0(980)$ mixing. Also, it is shown that the presence of the peak is not sensitive to the strong phases at short distance mentioned in Eq.~(9). If a peak for CP violation is observed in the experiments, it will be a strong evidence for the $a_0^0(980)$-$f_0(980)$ mixing.

At present, the $a_0^0(980)$-$f_0(980)$ mixing is examined by the mixing intensities in some isospin-violating decay processes. For example, the BESIII collaboration recently investigated the mixing intensities which are defined as
\begin{eqnarray}
   \xi_{fa}&=&\frac{B(J/\psi\rightarrow \phi f_0(980)\rightarrow \phi a^0_0(980)\rightarrow \phi\eta\pi)}{B(J/\psi\to\phi f_0(980) \to\phi\pi\pi)}, \\[6pt]
   \xi_{af}&=&\frac{B(\chi_{c1}\rightarrow \pi a^0_0(980)\rightarrow \pi f_0(980)\rightarrow \pi\pi\pi)}{B(\chi_{c1}\rightarrow \pi a^0_0(980)\rightarrow \pi\pi\pi))}.
\end{eqnarray}
In those studies, the mixing intensities were measured from the decay amplitudes. On the other hand, the CP asymmetry is proportional to sine of the strong phases which depends on the $a_0^0(980)$-$f_0(980)$ mixing more sensitively than the amplitudes. We propose to test the $a_0^0(980)$-$f_0(980)$ mixing by measuring the CP violating parameter. If the branching ratio for $B^\pm \to f_0(980) [a_0^0(980)] \pi^\pm$ is of the order $10^{-6}$ \cite{Tanabashi:2018oca}, then the number of $B \bar B$ pairs needed is roughly
 \begin{eqnarray}
N=\frac{n^2}{BR*A_{CP}^2}(1-A_{CP}^2)\nonumber,
\end{eqnarray}
which is $2\times 10^7$ for $3\sigma$ signature and $5\times 10^7$ for $5\sigma$ signature. With the increased data provided by LHC and forthcoming Belle-\uppercase\expandafter{\romannumeral2}, the study of CP violation in the $B$ meson has reached a high level of precision. The number of the $B \bar B$ pairs in LHCb could be around $10^{12}$ per year \cite{Bediaga:2018lhg} and the Belle detector has collected $8\times 10^8$ $B \bar B$ pairs \cite{Kaliyar:2018igf}, which are sufficient to check the $a_0^0(980)$-$f_0(980)$ mixing mechanism through CP violation.

Like the $\rho$-$\omega$ mixing, the $a_0^0(980)$-$f_0(980)$ mixing could occur in a lot of multi-body decay modes which contain the $S$-wave $\pi\pi$ final states in the $B$ or $D$ meson decays. We predict that the $a_0^0(980)$-$f_0(980)$ mixing would lead to large CP violation. Since the strong phase is more sensitive to the $a_0^0(980)$-$f_0(980)$ mixing than the decay widths, $A_{CP}$ is a more useful parameter to test the $a_0^0(980)$-$f_0(980)$ mixing. We calculate localized direct CP asymmetries in the three body decays $B^\pm \to f_0(980) [a_0^0(980)] \pi^\pm \to \pi^+ \pi^- \pi^\pm $ based on the hypothetical $q\bar q$ structure of $a^0_0(980)$ and $f_0(980)$ in the QCD factorization. It is shown that there is a peak for CP violation when the invariance mass of $\pi\pi$ is near 980\,MeV. This would be a new way to verify the $a^0_0(980)$-$f_0(980)$ mixing. Some input parameters are not well determined in our calculation and thus there are some uncertainties. One may also wonder if $a^0_0(980)$ and $f_0(980)$ produced in $B$ decays are dominated by the $q\bar q$ configuration and whatever the $a_0^0(980)$-$f_0(980)$ mixing could also provide a large strong phase and lead to large CP violation when $a^0_0(980)$ and $f_0(980)$ have other structures. This will need further investigation.

This work was supported by National Natural Science Foundation of China (Project Nos. 11805153, 11805012, 11575023 and 11775024), Natural Science Basic Research Plan in Shaanxi Province of China (Program No. 2018JQ1002).


\appendix
\section{}
In order to show the calculating process of the amplitudes clearly, we present explicit formulas for the vertex corrections $V_i(M_2)$, penguin corrections $P_i(M_2)$, hard spectator corrections $H_i(M_1,\,M_2)$, and $A^{i,\,f}_n$ in the weak annihilation contributions in the following.

The twist-2 light-cone distribution amplitude (LCDA) for the scalar meson $S$, $\Phi_S$, has the form \cite{Cheng:2005nb}
\begin{eqnarray}
 \Phi_M(x,\mu)=f_M6x(1-x)\left[1+\sum_{n=1}^\infty
 \alpha_n^M(\mu)C_n^{3/2}(2x-1)\right],
\end{eqnarray}
where $\alpha_n=\mu_SB_n$ with $B_n$ being the Gegenbauer moments and $C_n^{3/2}$ the Gegenbauer polynomials. For twist-3 LCDAs, we have $\Phi_S^s(x)=\bar f_S$ and $\Phi^\sigma_S(x)=\bar f_S6x(1-x)$. The twist-2 and twist-3 distribution amplitudes for the pseudoscalar meson $P$ are
\begin{eqnarray}
\Phi_P(x)= f_P6x(1-x),\qquad \Phi_P^p(x)= f_P,\qquad\Phi^\sigma_P(x)=f_P6x(1-x).
\end{eqnarray}

The expressions of the vertex corrections $V_i(M_2)$ are (apart from the decay constant $f_{M_2}$) \cite{Cheng:2005nb}
\begin{eqnarray}
 V_i(M_2) &=& 12\ln{\frac{m_b}{\mu}}-18-\frac{1}{2}-3\pi \mathrm i+\left(\frac{11}{2}-3\pi \mathrm i\right)\alpha_1^M-\frac{21}{20}\alpha_2^M+\left(\frac{79}{36}-\frac{2\pi \mathrm i}{3}\right)\alpha_3^M+\cdots, \nonumber \\
\end{eqnarray}
for $i=1-4,\,9,\,10$,
\begin{eqnarray}
 V_i(M_2) &=& -12\ln{\frac{m_b}{\mu}}+6-\frac{1}{2}-3\pi \mathrm i-\left(\frac{11}{2}-3\pi \mathrm i\right)\alpha_1^M-\frac{21}{20}\alpha_2^M-\left(\frac{79}{36}-\frac{2\pi \mathrm i}{3}\right)\alpha_3^M+\cdots, \nonumber \\
\end{eqnarray}
for $i=5,\,7$ and $V_i(M_2)=-6$ for $i=6,\,8$ in the naive dimensional regularization scheme for $\gamma_5$.

The hard spectator corrections read \cite{Cheng:2005nb}
\begin{eqnarray}
 H_i(M_1M_2) &=& {1\over f_{M_2}F_0^{BM_1}(0)m^2_B}\int^1_0 {\mathrm d\rho\over\rho}\Phi_B(\rho)\int^1_0 {\mathrm d\xi\over \bar\xi} \,\Phi_{M_2}(\xi)\int^1_0
{\mathrm d \eta\over\bar \eta}\left[\Phi_{M_1}(\eta)+r_\chi^{M_1}{\bar\xi\over\xi}\,\Phi_{m_1}(\eta)\right],\nonumber\\
\end{eqnarray}
for $i=1-4,\,9,\,10$,
\begin{eqnarray}
 H_i(M_1M_2) &=& -{1\over f_{M_2}F_0^{BM_1}(0)m^2_B}\int^1_0 {\mathrm d\rho\over\rho}\Phi_B(\rho)\int^1_0 {\mathrm d\xi\over \xi} \,\Phi_{M_2}(\xi)\int^1_0
{\mathrm d\eta\over \bar\eta}\left[\Phi_{M_1}(\eta)+r_\chi^{M_1}{\xi\over\bar\xi}\,\Phi_{m_1}(\eta)\right],\nonumber\\
\end{eqnarray}
for $i=5,\,7$ and $H_i=0$ for $i=6,\,8$, where $\bar\xi\equiv 1-\xi$
and $\bar\eta\equiv 1-\eta$, $\Phi_M$ ($\Phi_m$) is the twist-2 (twist-3) light-cone distribution amplitude of the meson $M$.

The explicit expressions of $A^{i,\,f}_n$ in the weak annihilations are given by \cite{Cheng:2005nb}
\begin{eqnarray}
 A_1^{i}&=& \int\cdots \begin{cases} \left( \Phi_{M_2}(x)\Phi_{M_1}(y)\left[{1\over y(1-x\bar y)}+{1\over \bar x^2y}\right]-r_\chi^{M_1}r_\chi^{M_2}\Phi_{m_2}(x)\Phi_{m_1}(y)\,{2\over \bar xy}\right); & \mathrm{for}~M_1M_2=PS,  \cr
 \left( \Phi_{M_2}(x)\Phi_{M_1}(y)\left[{1\over y(1-x\bar y)}+{1\over \bar x^2y}\right]+r_\chi^{M_1}r_\chi^{M_2}\Phi_{m_2}(x)\Phi_{m_1}(y)\,{2\over \bar xy}\right); & \mathrm{for}~M_1M_2=SP, \end{cases} \nonumber  \\[5pt]
 A_2^{i}&=& \int\cdots \begin{cases} \left( -\Phi_{M_2}(x)\Phi_{M_1}(y)\left[{1\over \bar x(1-x\bar y)}+{1\over \bar xy^2}\right]+r_\chi^{M_1}r_\chi^{M_2}\Phi_{m_2}(x)\Phi_{m_1}(y)\,{2\over \bar xy}\right); & \mathrm{for}~M_1M_2=PS,  \cr
 \left( -\Phi_{M_2}(x)\Phi_{M_1}(y)\left[{1\over \bar x(1-x\bar y)}+{1\over \bar xy^2}\right]-r_\chi^{M_1}r_\chi^{M_2}\Phi_{m_2}(x)\Phi_{m_1}(y)\,{2\over \bar xy}\right); & \mathrm{for}~M_1M_2=SP, \end{cases}  \nonumber \\[5pt]
 A_3^i&=& \int\cdots \begin{cases} \left( r_\chi^{M_1}\Phi_{M_2}(x)\Phi_{m_1}(y) \,{2\bar y\over \bar xy(1-x\bar y)}+r_\chi^{M_2}\Phi_{M_1}(y)\Phi_{m_2} (x)\,{2x\over \bar xy(1-x\bar y)}\right); & \mathrm{for}~M_1M_2=PS,  \cr
 \left( -r_\chi^{M_1}\Phi_{M_2}(x)\Phi_{m_1}(y) \,{2\bar y\over \bar xy(1-x\bar y)}+r_\chi^{M_2}\Phi_{M_1}(y)\Phi_{m_2}(x)\,{2x\over \bar xy(1-x\bar y)}\right); &  \mathrm{for}~M_1M_2=SP, \end{cases} \nonumber \\[5pt]
 A_3^f &=& \int\cdots \begin{cases} \left(r_\chi^{M_1} \Phi_{M_2}(x)\Phi_{m_1}(y)\,{2(1+\bar x)\over \bar x^2y}-r_\chi^{M_2} \Phi_{M_1}(y)\Phi_{m_2}(x)\,{2(1+y)\over \bar xy^2}\right); & \mathrm{for}~M_1M_2=PS,\cr
 \left( -r_\chi^{M_1} \Phi_{M_2}(x)\Phi_{m_1}(y)\,{2(1+\bar x)\over \bar x^2y}-r_\chi^{M_2} \Phi_{M_1}(y)\Phi_{m_2}(x)\,{2(1+y)\over \bar xy^2}\right); & \mathrm{for}~M_1M_2=SP,\end{cases} \nonumber \\[7pt]
  A_1^f &=& A_2^f=0,
\end{eqnarray}
where $\int\cdots=\pi \alpha_s\int^1_0\mathrm d x\mathrm d y$, $\bar x\equiv1-x$ and $\bar y\equiv1-y$.

\end{document}